
\overfullrule=0pt
\magnification=1000
\vsize=21.0 truecm
\hsize=14.5 truecm
\baselineskip=12pt
\hoffset=1.5 truecm
\voffset=1.0 truecm

\font\ftext=cmr10
\font\ftit=cmbx10
\font\abstract=cmr10

\parskip=6pt
\parindent=2pc

\font\titulo=cmbx10 scaled\magstep1

\def\ii{\'\i}

\def\section#1{\vskip 0.5truepc plus 0.1truepc minus 0.1truepc
	\goodbreak \leftline{\titulo#1} \nobreak \vskip 0.1truepc
	\indent}
\def\frc#1#2{\leavevmode\kern.1em
	\raise.5ex\hbox{\the\scriptfont0 $ #1 $}\kern-.1em
	/\kern-.15em\lower.25ex\hbox{\the\scriptfont0 $ #2 $}}

\def\Real{{\rm I\!R}}  

\rightline{ICN-UNAM, Mexico, January 20, 1994.}

\vskip 2.5pc

\centerline{\ftit CANONICAL QUANTIZATION OF (2+1)-DIMENSIONAL
GRAVITY \footnote{*}{
\ftext{This work is supported in part by the National Science
Foundation, Grant No. PHY89-04035, by CONACyT grant No.
400349-5-1714E and by the Association G\'en\'erale pour la
Coop\'eration et le D\'eveloppement (Belgium).}}}

\vskip 1.5pc

\centerline{Henri Waelbroeck}

\vskip 0.5pc

\centerline{Institute for Theoretical Physics}
\centerline{University of California at Santa Barbara}
\centerline{Santa Barbara, CA 93106-4030}

\vskip 0.2pc

\centerline{and}
\centerline{Instituto de Ciencias Nucleares, UNAM}
\centerline{Apdo. Postal 70-543, M\'exico, D.F., 04510 M\'exico}

\vskip 1.2pc

\centerline { Abstract}

{\leftskip=2.5pc\rightskip=2.5pc\smallskip\noindent We consider
the quantum dynamics of both open and closed two-dimensional
universes with ``wormholes'' and particles.  The wave function is
given as a sum of freely propagating amplitudes, emitted from a
network of mapping class images of the initial state.
Interference between these amplitudes gives non-trivial scattering
effects, formally analogous to the optical diffraction by a
multidimensional grating; the ``bright lines'' correspond
to the most probable geometries.

\smallskip}

\smallskip
\noindent

\vfill
\eject

\section{1. Introduction}

	Almost thirty years after the first serious attempts at formulating
a quantum theory of gravity [1], still very little is known about
what such a theory might be, beyond the semiclassical approximation.
It is non-renormalizable as a quantum field theory, and there is a
general feeling that one must seek a non-perturbative formulation
of quantum gravity, since the basic concept of a smooth geometry
becomes incompatible with Heisenberg's uncertainty principle at
very small scales.

	Unfortunately, the non-perturbative analysis of nonlinear
dynamical systems is a notoriously difficult problem.  Quantum
gravity is far from the only case where non-perturbative effects
are important.  Other examples include some
of the most pressing problems at hand, such as climatology and
plasma physics.  Certainly, the challenge of developping new methods
for this class of problems must be accepted.  Theoretical
work in the field has split along two currently non-intersecting
trajectories; one deals with chaos, the other is related to knots
(or more generally, things that cannot be made to vanish by a
succession of small deformations).  Although both are studied in
the context of gravity, knots appear to be most closely related to
quantum gravity and the issue of the ``small scale structure'' of
spacetime: relations between quantum gravity and knots have
appeared both in $3+1$ dimensions [2], and in the model of
$(2+1)$-dimensional gravity [3], which we will pursue in this article.

	Before we begin, it should be stressed that there is still no
candidate for a consistent theory of ``quantum gravity", nor a
proof that such a theory exists, so in a way our problem is
worse than, say, climatology.  On the other hand, numerous
authors have suggested that $(2+1)$-dimensional gravity could be
used as a simpler model, in which one could  formulate a quantum
theory and investigate issues of interpretation, the choice of
time, etc. [4].

	Witten's work on the subject has led to a complete set of
Heisenberg picture observables for the quantum theory, an
important step in the quantization programme [5].  In spite of this,
much remains to be done before one could say that quantum gravity
in 2+1 dimensions has been solved.  The authors who have undertaken
to follow up Witten's work have encountered major obstacles in
two different directions.  Nelson and Regge, trying to make Witten's
reduced phase space explicit, found that they could do so up to
genus two only by finding the ideal of a complicated system of
relations among traces, some of which are cousins of the
Cayley-Hamilton relations for SO(3) matrices [6].  On the other
hand, the task of extracting the explicit time dependence of the
observables (i.e., solving the Heisenberg equations for some
choice of internal time) was solved in the case of genus one
universes by Moncrief [7], and in the context of Witten's
variables, by Carlip [8].  Interestingly, these authors differred
as to what the Hamiltonian should be, in particular Carlip's
proposed Hamiltonian was unbounded from below.  Following
Moncrief's remarks to this effect, Carlip showed that his
formulation led to a ``Dirac square root" of the Wheeler-DeWitt
equation, when one demands invariance of wave functions with
respect to the mapping class group [9].

	Our purpose in this article is to pursue the quantization of
$(2+1)$-dimensional gravity to the point of providing a complete
set of Schroedinger observables, a Hamiltonian, a consistent
choice of ordering of the operators and, in computable form, wave
functions for various types of scattering problems.  We hope that
this will help to formulate some of the fundamental problems
of quantum gravity in terms of specific,
computable questions.  We stress that some of these ``problems"
involve the very consistency of the formalism which we are
proposing, so we do not claim to have proved that the theory exists.

	Many authors have worked on $(2+1)$-dimensional quantum gravity,
and it sometimes seems that each one has a different approach.
We will briefly list some of these approaches in terms of two
main schools of thought.  The issue is how to reduce the infinite
dimensional phase space by exploiting the infinite number of
gauge symmetries, to obtain a finite-dimensional system which
could be quantized exactly.  One approach is to fix the gauge
by choosing a slicing of spacetime in constant curvature slices
(York gauge [10]), or some other choice of slicing - we will
call this the ADM school.  The other relies on the fact that
$(2+1)$-dimensional spacetimes can be regarded as a particular
sheet of solutions (with ``maximal Euler class") of the Chern-
Simons gauge theory for an $ISO(2,1)$ connection - the $ISO(2,1)$
school.

\noindent 1- The ADM School

\noindent 1.1 Particles and cones.

 The geometry of an $\Real ^{2}$ universe with particles is a
multi-cone, i.e. a flat surface with conical singularities, where
the deficit angle at each singularity is related to the mass of the

particle.  The two-particle classical dynamics was solved explicitly
by Deser, Jackiw and 't Hooft [11], the quantum scattering of a light
particle by a heavy one was worked out by Deser and Jackiw [12],
and the general two-particle quantum scattering problem
was solved by 't Hooft [13].  The three-particle classical problem
was solved by Lancaster and Sasakura [14].

\vskip 0.2pc
\noindent 1.2  Riemann surfaces.

 The geometry is a genus $g$ Riemann surface, from a constant extrinsic
curvature foliation of spacetime.  The surface coordinates are chosen
so that the metric is conformally flat (Hosoya and Nakao) or
conformally hyperbolic (Moncrief).  To compute the Hamiltonian,
one solves an equation for the conformal factor.  In the first
case, it is not a coordinate scalar so the equation must be solved
on coordinate patches - it is not known whether it admits a
solution for genus $g > 1$.  In the conformally hyperbolic case,
Moncrief finds a Lichnerowitz equation for the scalar conformal
factor, and proves the existence of a solution for all $g$ [7].
The Shroedinger equation is easily written down for genus one [15],
and the solutions can in principle be found from the modular
invariant eigenfunctions of the Laplacian on Teichmuller space,
which are known as the Maass forms [16].

 The virtue of the ADM approach is that it is most similar to
$(3+1)$-dimensional gravity.  Its drawback is that it encounters
some technical problems that are most similar to those of
$(3+1)$-dimensional gravity.

\vskip 0.5pc

\noindent 2- The ISO(2,1) School

\noindent 2.1 Reduced phase space quantization.

 One finds a complete set of phase space observables which commute
with the constraints.  The reduced phase space is related to the
moduli space of flat $ISO(2,1)$ connections (Witten, [5]).  The problems
begin when one attempts to parametrize this space and compute the
Poisson brackets: one can form an infinite
set of $ISO(2,1)$ invariants from the holonomy, but these
are not independent variables:  They are related by non-linear
constraints, some of which are cousins of the Cayley Hamilton
identities.  Nelson, Regge, Urrutia and Zertuche [17] have succeded
in identifying the reduced phase space explicitly up to genus $g=2$.
The canonical quantization process leads, in some cases, to quantum
groups [18].  The reduced phase space can be interpreted as a
Hamilton-Jacobi formulation [19], so the representations of the
reduced algebra should give the wave functions in the Heisenberg
picture. No attempt is made to compute the Hamiltonian or to
formulate quantum dynamics questions.

\noindent 2.2 Covariant quantization.

 One works with the $ISO(2,1)$ ``homotopies" along loops which form an
arbitrarily chosen basis of $\pi_1(\Sigma_g)$, and later
imposes invariance with respect to a change of basis, or mapping
class transformation (Carlip, [20]).  The difficulties encountered by
Regge et al. can be avoided by using these homotopies rather than the
invariants derived from them, leaving out a global $ISO(2,1)$ symmetry.
The homotopies can be parametrized in the ``polygon
representation" of $(2+1)$-dimensional gravity, and the symplectic
structure is known (H.W., [21]).  Mapping class invariant scattering
amplitudes can be computed by the method of images, by summing
over all mapping class images of the 'in' state.  Two problems
are encountered:  Each term in the sum can only be computed if
one knows the Hamiltonian, for a given choice of time.  The maximal
slicing choice only allows one to compute the Hamiltonian explicitly
only for genus $g = 1$ (see ADM school).  The other problem
is that the sum is over the non-abelian, infinite, mapping class
group, and it is not clear a priori which terms might be small, or
if the sum converges.

\vskip 2pc
\noindent 2.3 Planar Multiple Polygons

 Recently, 't Hooft has proposed another approach based on a
slightly different polygon representation, where one chooses a
piecewise flat surface, which leads to a number of planar polygons,
with $SO(2,1)$ identifications of the boundary edges. There is a
remarkably simple symplectic structure and Hamiltonian for this
system; since these variables are directly related to a choice of
generators of the first homotopy group, the mapping class group
acts non-trivially on the wave functions [22].

	This article can now be summarized in one line:  We combine our
earlier calculation of the Hamiltonian in the polygon representation,
with the method of images and the ``stationnary phase theorem", to
advance along the covariant quantization programme [23].

	The reader is referred to the previous article on the polygon
representation of $(2+1)$-dimensional gravity for the canonical variables,
the Hamiltonian, and most importantly, the relation between the
polygon representation and the $ISO(2,1)$ homotopies [19].  We will
begin this article directly with the canonical quantization in
Sec. 2.  The mapping class group invariance is discussed in
Sec. 3, where we write down invariant amplitudes for various types
of scattering problems. In Sec. 4 we give the conclusions and discuss
the problem of time.

\vfill
\eject

\section{2. Quantization in the Polygon Representation}

	In this section, we will draw heavily on results from the
classical theory; their derivations can be found in [19] and
references therein, and will not be
repeated here.  The reduced phase space of $(2+1)$-dimensional
gravity can be parametrized by $2g+N$ three-vectors $E(\mu )$ and
as many $SO(2,1)$ matrices $M(\mu )$, with the following Poisson
brackets and constraints $(\mu =1,2,..,2g+N)$.

$$\{ E^a (\mu), E^b (\mu)\} = \epsilon^{abc} E_c (\mu) \eqno(2.1)$$

$$\{ E^a(\mu), M^b{}_c(\mu)\} = \epsilon^{abd} M_{dc}(\mu) \eqno(2.2)$$

$$\{ M^a_{\ b}(\mu), M^c_{\ d}(\mu) \} = 0 \eqno(2.3)$$

\noindent The vectors $E^a(\mu )$ together with their identified
partners $-M^{-1}(\mu ) E(\mu )$ form a closed polygon,

$$J \equiv \sum (I - M^{-1}(\mu)) E(\mu) \approx 0 \eqno	(2.4)$$

\noindent and the cycle condition for the $SO(2,1)$ identification
matrices leads to the constraints

$$P^a \equiv {1\over 2} \epsilon^{abc} W_{cb} \approx 0 \eqno(2.5)$$

\noindent where

$$\eqalign{& W = \biggl( M(1) M^{-1}(2) M^{-1}(1) M(2)\biggr)
\cdots \biggl( M(2g-1) M^{-1}(2g) \cr
& M^{-1}(2g-1) M(2g)\biggr) M(2g+1) M(2g+2) \cdots
M(2g+N)\cr} \eqno(2.6)$$

The constraints $J^a$ and $P^a$ generate a Poincar\' e algebra
with the brackets (2.1)-(2.3).  The masses of the $N$ particles
are given by the constraints

$$H(\mu) \equiv P^2(\mu) + sin^2 \left( \Omega (\mu)\right) \approx 0
\eqno(2.7)$$

\noindent which generate translations of each particle along
its worldline.

It is convenient to define the canonical variables $P^a(\mu )
= (1/2) \epsilon^{abc} M_{cb}(\mu )$ and

$$X(\mu) = {1\over P^2(\mu)} \biggl( P(\mu) \wedge J(\mu) - {2(E(\mu)
\cdot P(\mu)) P(\mu) \over tr M(\mu)-1} \biggr) \eqno(2.8)$$

\noindent where

$$J(\mu) = (I-M^{-1}(\mu)) E(\mu)\eqno(2.9)$$

\noindent The variables $X(\mu )$ and $P(\mu )$ have canonical
Poisson brackets:

$$\{ P_a(\mu), \ X^b(\mu) \} = \delta_a{}^b \eqno(2.10)$$

The constraints can be written in terms of the canonical
variables, using $J(\mu ) = X(\mu ) \wedge P(\mu )$ and

$$M^a{}_b = \delta^a{}_b + P_c \epsilon^{ca}{}{}_b + \biggl(
\sqrt{1+P^2}-1 \biggr) \biggl( \delta^a{}_b - {P^a P_b\over P^2}
\biggr) \eqno(2.11)$$

\noindent for $\mu =1,...,2g$ (for which $M(\mu )$ is hyperbolic
(a boost) [24]), or for $\mu =2g+1,...,2g+N$,

$$M^a{}_b = \delta^a{}_b + P_c \epsilon^{ca}{}{}_b + \biggl(
\sqrt{1-P^2} -1 \biggr) \biggl(\delta^a{}_b - {P^a P_b\over P^2}
\biggr) \eqno(2.12)$$

\noindent The $SO(2,1)$ constraints become

$$J \equiv \sum_{\mu} X(\mu) \wedge P(\mu) \approx 0 \eqno (2.13)$$

\noindent while the translation constraints $P \approx 0$ are
defined implicitly in terms of ${P(\mu )}$ by (2.11) and (2.12).
The explicit form of the function $P(P(\mu )) \approx 0$
can be derived from the relations (2.11)-(2.12).

	We will assume the universe is a closed surface
with genus $g > 1$; for the other cases the arguments below carry
over substitutes for the ``time" and ``Hamiltonian", the
appropriate expressions drawn from the previous article.
We propose the following choice of ``internal time", in a frame where
$M(1)$ is a boost in the $(y \ t)$-plane:

$$T = - {E^x (1) \over P^x (1)} \eqno(2.14)$$

\noindent The Hamiltonian is given in terms of the variables$ M(\mu ),
\mu > 1$, by solving the constraint

$$W \equiv \biggl( M(1) M^{-1}(2)M^{-1}(1)M(2)\biggr) \cdots \biggl(
M(2g-1) M^{-1}(2g)M^{-1}(2g-1)M(2g)\biggr) \approx I \eqno(2.15)$$

\noindent for $tr(M(1))$, where $M(1)$ is a boost of magnitude $b$
with axis $x$. One finds

$$H = {P_t(2) Q_t - P_y(2) Q_y \over P_y^2(2) - P_t^2(2)} \eqno(2.16)$$

\noindent where

$$\eqalign{ Q^a & = {1 \over 2} \epsilon^{abc} \biggl( M(2) (M(3)
M^{-1}(4)M^{-1}(3)M(4)) \cdots \cr
& (M(2g-1) M^{-1}(2g) M^{-1}(2g-1) M(2g))\biggr)_{cb}\cr}\eqno(2.17)$$

There are two residual ``translation" constraints and one ``boost"
constraint. The former come from $M(1) M^{-1}(2) M^{-1}(1)$
$\approx (M(2) M(3) M^{-1}(4) M^{-1}(3) M(4) \cdots M(2g+N))^{-1}$,

$$P^x (2) - Q^x \approx 0 \eqno(2.18)$$

$$P^2 (2) - Q^2 \approx 0 \eqno(2.19)$$

\noindent The boost generator is

$$J^x \equiv \sum_{\mu > 1} \biggl( I-M^{-1}(\mu) \biggr)^x_{\ a}  \ E^a(\mu)
\approx 0 \eqno(2.20)$$

\noindent The remaining phase space variables are $E(\mu )$,
$M(\mu )$ for $\mu = 2, ..., 2g+N$.  Their time evolution is
generated by the Hamiltonian (2.16) and any linear combination
of the first-class constraints (2.18)-(2.20), which commute with
the Hamiltonian.

If $N \not= 0$, one can choose a gauge where the vector $X(2g+1)$
is aligned with the $x$-axis and $X(2g+2)$ vanishes at all times;
$M(2g+2)$ is then a pure rotation with angle $\Omega (2g+2)$, and

$$T = {X^x (2g+1) \over P^x (2g+1)} \eqno(2.21)$$

$$P^x \equiv \epsilon^{xbc} \biggl( M(2g+3) \cdots M(2g+N) M(1) \cdots
M(2g) M(2g+1) \biggr) ^{-1}_{\ \ cb} \approx 0 \eqno(2.22) $$

$$P^y \equiv \epsilon^{ybc} \biggl( M(2g+3) \cdots M(2g+N) M(1) \cdots
M(2g) M(2g+1) \biggr) ^{-1}_{\ \ cb} \approx 0 \eqno(2.23) $$

$$H = Tr \biggl( M(2g+2) \cdots M(2g+N) M(1) M^{-1}(2) \cdots
M^{-1}(2g-1) M(2g) \biggr) \eqno(2.24)$$

Note that the internal times (2.14) and (2.21) are
linearly related, as well as any choice which is linear in the E's,
since $d^2 E(\mu ) /dt = 0$.  Such times are also ``inertial",
in the sense that the position of a particle as seen from the
observer, $E(\mu )$, follows a straight trajectory at constant
``velocity".

	We begin by constructing the quantum theory in the most naive
way, then pick up the problems as they appear (see the mapping
class group, next section).  We quantize the canonical phase
space $X(\mu ), P(\mu )$ in the usual way ($\mu = 2,3,...,2g+N$).

$$P_a(\mu) = -i {\partial \over \partial X^a(\mu)} \eqno(2.25)$$

$$[P_a(\mu), \ X^b(\mu)] = -i \delta_a{}^b \eqno(2.26)$$

	To prove the equivalence of this quantization to one which could
be carried out in the variables {$E(\mu ), M(\mu )$}, one must express
these variables in terms of $X(\mu )$ and $P(\mu )$,
and show that there exists an ordering of these expressions
such that the commutators form a representation of the algebra
(2.1)-(2.3).  An acceptable ordering is obtained by placing all
the $X$'s on the right:

$$E(\mu) = {M^{1/2} \over P \sqrt{2\sqrt{P^2+1}-2}} P \wedge
(P\wedge X) - {\sqrt{P^2+1} \over P^2} P (P \cdot X)
\eqno(2.27)$$

\noindent where the index $\mu =2,...,2g+N$ has been omitted in the right
hand side expressions, and $M^{1/2}$, the ``square root" of $M(\mu)$,
was given explicitly in the previous article, Eqns. (6.7) and (6.9).

	As mentioned, the ordering of the operators in (2.27) leads to the
correct commutator algebra for $E(\mu ), M(\mu )$.  For example,
using $P^a (\mu ) = -i \eta^{ab} \partial / \partial X^b(\mu )$,
one finds $[E^a(\mu ), E^b(\mu )] = -i \epsilon^{abc}E_c(\mu )$.
This solution to the ordering problem is not unique.

	The Wheeler DeWitt quantization would consist in taking the set
of square integrable functions on $\Real^{3 \times (2g+N)}$, which
are annihilated by the constraint operators:

$$\psi = \psi(\{ X(\mu); \mu = 1, \cdots, 2g+N \} ) \eqno(2.28)$$

$$J^a \psi = 0 \eqno(2.29)$$

$$P^a \psi = 0 \eqno(2.30)$$

\noindent Equations (2.29) state that $\psi$ must be a Lorentz scalar in
the variables $X(\mu )$.  The other constraints are the Wheeler-
DeWitt equations.  These can in principle be obtained explicitly
from (2.11), (2.12) by replacing everywhere $P_a(\mu )$ by the
operator $-i \hbar \partial / \partial X^a(\mu )$.  The result is
a non-rational function of the derivatives, not likely to give
a physically acceptable quantum theory.  For this
reason, we will prefer to partly fix the gauge as indicated above,
by choosing the internal time (2.21) and the corresponding
Hamiltonian (2.24), and quantize in the Schroedinger picture:

$$i \hbar {\partial \psi \over \partial T} = H(\{ P(\mu); \mu =
2, \cdots, 2g+N \}) \psi \eqno(2.31)$$

One must still impose the two ``momentum constraints" (2.22)-(2.23),

$$\epsilon^{xbc} \biggl( M(2g+3) \cdots M(2g+N) M(1) \cdots
M(2g) M(2g+1) \biggr) ^{-1}_{\ \ cb} \ \ \psi = 0 \eqno(2.32)$$

$$\epsilon^{ybc} \biggl( M(2g+3) \cdots M(2g+N) M(1) \cdots
M(2g) M(2g+1) \biggr) ^{-1}_{\ \ cb} \ \ \psi = 0 \eqno(2.33)$$

\noindent and require that the wave function be a Lorentz scalar.

The solution to the Schroedinger equation, for an initial state
$\psi ({X(\mu )}, 0)$, $\mu = 2, 3,..., 2g+N$, is given by

$$\psi( \{ X(\mu) \}; \ T) = e^{i T H(\{ -i {\partial / \partial
X(\mu)} \} } \ \ \psi( \{ X(\mu) \}; \ 0) \eqno(2.34)$$

The Hamiltonian operator is defined by the Fourier decomposition:

$$e^{i T H(\{ -i {\partial / \partial X(\mu)} \} } \
e^{i K_0 \cdot X} = e^{i T H(K_0)} \eqno(2.35)$$

	Finding initial states which satisfy the constraints is not
difficult in principle.  Only the constraints (2.32) and (2.33),
which depend on the momenta in a non-polynomial way, appear to
present a difficulty, but in $P$-space these become algebraic
relations, which summarize the cycle conditions, for which
explicit solutions can be constructed (this is not always easy).

\vfill
\eject

\section{3. Mapping Class Group and Quantum Scattering}

	The preliminary theory of $(2+1)$-dimensional quantum gravity
which we constructed in Sec. 2, is formally similar to the
non-relativistic quantum mechanics of a set of free particles
propagating in a flat three-dimensional background, but with an
unusual ``kinetic energy" function.  This non-interacting picture
can be understood as follows:  We have defined the polygon
variables by cutting up and unfolding the manifold, to get a
polygon embedded in Minkowski space.  In this classical picture,
the corners of the polygon follow straight timelike lines - so
indeed they behave very much like free particles!

\

\noindent 3.1 The Mapping Class Group

	The variables which we are using are associated to a particular
choice of generators of the fundamental group for a genus $g$
surface with $N$ punctures.  A different choice of generators
can lead to a different set of variables for the same spacetime.
The group of transformations from one set of generators to another,
called ``mapping class group", reflects a discrete symmetry
of the polygon representation. One must demand that the wave function
be mapping-class invariant, which leads to interesting ``scattering"
effects [23].

	Before we begin to construct invariant wave functions, we will
review the mapping class group and its action on the polygon
variables, including ``time" and the Hamiltonian. We must also
make sure that our choice of operator ordering is consistent with
this symmetry group.

	The mapping class group is presented by the following generators
[25] (the loops $a_i, b_i$ are a standard basis, where each $a_i$
intersects only the corresponding $b_i$ at only one point;
$\tau_{u_i}, \tau_{y_i}, \tau_{z_i}$ are Dehn twists [20]).

\

$$\tau_{ui} : b_i \to b_i a_i \eqno(3.1)$$

$$\tau_y: a_i \to a_i b_i^{-1} \eqno(3.2)$$

$$\tau_{zi}: \cases{ a_i \to a_i b_i^{-1} a_{i+1} b_{i+1} a_{i+1}^{-1}
\hfill \qquad\qquad\qquad\qquad(3.3)\cr
b_i \to a_{i+1} b_{i+1}^{-1} a_{i+1}^{-1} b_i a_{i+1} b_{i+1} a_{i+1}^{-1}
\hfill \qquad\qquad\qquad\qquad (3.4)\cr
a_{i+1} \to a_{i+1} b_{i+1}^{-1} a_{i+1}^{-1} b_i a_{i+1} \hfill
\qquad\qquad\qquad\qquad (3.5)\cr}$$

 For genus $g = 2$, the representation of these generators in
$ISO(2,1)$ was given in [19], and can be recovered with the change
in notation ${u_i \to a_i , v_i \to b_i}$.  For example,

$$\rho (a_1) = \pmatrix{M(1) & (I-M(1)) E(1) - M(1)E(2) \cr
& + (I-M(1)) OA\cr \ & \ \cr
0 & \qquad \qquad 1 \cr} \eqno(3.6)$$

$$\rho(b_1) = \pmatrix{ M^{-1}(2) & (I-M^{-1}(2) - M^{-1}(1)) E(1)
\cr
& +  (I-M^{-1}(2)) E(2) + (I-M^{-1}(2))OA\cr \ & \ \cr
0 & \qquad \qquad 1 \cr} \eqno(3.7a)$$

\noindent so that we find, applying Equation (3.1), that $\rho (a_1)$
is unchanged and $\rho (b_1)$ becomes

$$\rho(b_1) \to \pmatrix{ M^{-1}(2) M(1) & (I-M^{-1}(1) - M^{-1}(2)
M(1)) E(1)\cr
& +(I-M^{-1}(2) - M^{-1}(2)M(1)) E(2)\cr
& + (I-M^{-1}(2)M(1)) OA \cr \ & \ \cr
0 & \qquad\qquad 1\cr} \eqno(3.7b)$$

\noindent By comparing this to the expression (3.7a), one can deduce
the action of the mapping class transformation on the
variables $E(\mu )$ and $M(\mu )$ (this is not easy).

$$\tau_{u_i}: \cases{ E(1) \to E(1) + E(2) \hfill(3.8)\cr
E(2) \to M^{-1}(1) E(2) \hfill \qquad\qquad\qquad\qquad (3.9)\cr
M(1) \to M(1)  \hfill \qquad\qquad\qquad\qquad (3.10)\cr
M(2) \to M^{-1}(1) M(2) \hfill \qquad \qquad \qquad \qquad (3.11)\cr}$$

\noindent and similarly for $\mu  = 3$ and $\mu = 4$, and for any other
element of the mapping class group.  These relations can also be
derived from the polygon representation of the variables
$E(\mu )$ and $M(\mu )$ (Figure 3.1).  The loop $a_1$ is a loop
starting at $O$, going through the segment $E(1)$ at the point
$X$, through $M^{-1}(1)E(1)$
at $X'$ and back to $O$.  Likewise $b_1$ starts at $O$,
goes through the segment $M^{-1}(2) E(2)$ at $Y'$, then
continues through $Y$ and returns to $O$.  Now one can deform
the combined loop $b_1 a_1$  smoothly into the loop which is
represented in Figure 3.1, where a small semicircle surrounds
the corner $B$ of the polygon.  One can then choose a new set
of cuts which avoids the corner $B$:  The first cut is now
$E(1) + E(2)$ (dashed line), while the second cut is as before
but in a frame which has been transported around the loop $E(2)$
once: $E(2) \to M^{-1}(1) E(2)$.  Any mapping class group
transformation can be obtained in a similar fashion from the
polygon representation, by sliding one corner of the polygon
around a loop (a Dehn twist).

	Mapping class transformations leave the cycle conditions
invariant, since they are automorphisms of $\pi_1(\Sigma)$.  This
implies that the Hamiltonian, which was computed in the previous
section by solving the cycle conditions in a frame where $M(1)$
is a pure $x$-boost, is invariant as long as the frame is
adjusted so that this condition is preserved.  We will thus choose
the frame independently for each term in the sum over the mapping
class group. This requires adjusting the initial and final
conditions appropriately, by an overall $SO(2,1)$ transformation.
Time, defined as the variable canonically conjugate to this Hamiltonian,
is mapping-class invariant up to a constant.

	One must check that the mapping class group action is consistent
with the choice of operator ordering.  If one chooses to place
the $E$'s and $X$'s on the right, the bracket algebra, as well
as the relations that give $X(\mu )$ as a function of $E(\mu )$,
are preserved under the mapping class group transformations
(this can be checked explicitly, and derives from the fact that all
the relations are linear and homogeneous in $E$ and $X$).  This
differs significantly from the situation in York's extrinsic gauge,
where the task of finding an operator ordering consistent with
the mapping class group generally requires that one consider
multi-valued wave functions, with a modular weight which depends
on the ordering chosen [9].  We feel that this requirement is due
to the choice of slicing, and is not a fundamental property of
$(2+1)$-dimensional gravity.  Indeed, the polygon variables do
not exclude the possibility of choosing the extrinsic curvature
slicing; rather they do not specify any foliation:  We have
decoupled the slicing (gauge) from the global variables
(observable), and simply dropped the former.  The difficulty
discovered by Carlip, is that it is difficult to find a
consistent operator ordering for the relations which give
the ADM operators in terms of the quantized $ISO(2,1)$ homotopies.

\

\noindent 3.2 Invariant wave functions - The method of images

	One must find mapping class invariant wave functions which
solve the Schroedinger equation.  We first review what has
already been achieved in both the ADM and the $ISO(2,1)$ schools.
In the ADM approach [7, 26], the Hilbert space is the space of
modular invariant square integrable functions over Teichmuller
space.  To find the solutions to the Schroedinger equation,
one considers the eigenfunctions of the Hamiltonian operator.  The
Hamiltonian is not yet known for genus greater than one
(one must solve a Lichnerowitz equation), but it is expected
that if a solution could be found, then the ordering problems
would be intractable.  This means that one is probably limited
to the case of the torus, where the Hamiltonian is known and
efforts are under way to compute the wave functions explicitly.

	In the $ISO(2,1)$ approach, one computes the amplitude for
the scattering $N$ particles on a plane as a sum of
amplitudes over all possible braidings of the worldlines of
the particles; if one is dealing with a compact universe with
punctures, then one sums over the mapping class group, rather
than the braid group.  Each term in this sum is a path integral
with the Chern-Simons action, over all Poincar\' e connections
on three-manifolds with fixed ``initial" and ``final" conditions,
and a given topology and mapping (or ``braiding", for particles).
Besides the difficulty of computing the path integral (which
is related to the difficulty of finding the Hamiltonian),
one must deal with an infinite sum over a non-abelian group;
it is not known whether the sum can be ordered as a converging
series, or if the result is computable.

	In both approaches, one can separate the task at hand into
two sub-tasks:

\noindent (1)  The dynamical problem (finding the Hamiltonian,
or computing a sum over histories with fixed mapping).

\noindent (2)  The mapping class problem (finding modular-
invariant eigenstates of the Hamiltonian operator, or performing
the sum over the mapping class group).

	Both the dynamical and the mapping class problems
have been solved for the scattering of two particles on $\Real^2$
[13], and the Hamiltonian is known for the quantum torus [7], with
work under way on the mapping class problem [33].

	So far in this article, we have solved the dynamical problem
in the general case of a genus g surface with N punctures.  In
what follows, we will attempt to provide a solution to the
mapping class problem, by constructing the invariant wave
functions as a series of computable terms.  We will give a
procedure to decide on which terms (or elements of the mapping
class group) give a significant contribution to the amplitude,
and argue that the number of such terms is finite.

	We will consider the amplitude to go from an initial state
$\mid X_1 >$, which we choose to be an eigenstate of the ``position"
operators at $T_1 = 0$, $\{ X^a(\mu ); \mu = 2,\cdots ,2g+N \}$, to a
final state  $\mid X_2 >$ at $T_2 > 0$.  Neither of these states
is mapping class invariant, but one can construct an invariant
states $\mid X >_{inv}$ by summing over all mapping class images
of X:

$$\vert X >_{inv} \sim \sum_{i \epsilon\{m.c.g\}} \vert \rho(g_i) X
> \eqno(3.12)$$

This eignestate has infinite norm, equal to the cardinality of
the mapping class group. We will
use the formal expression (3.12) as a starting point, and derive
physically acceptable wave functions below.

\

\noindent 3.3 Scattering Amplitudes

	Even though the spacetimes considered here are not asymptotically
flat, one can define non-interacting ``in" and ``out" states as
follows.  Since there is no potential energy term in the
Schroedinger equation, the system is non-interacting as long
as the mapping class group symmetry is not taken into account.
It is easy to see from the construction of invariant states in
(3.12), how the mapping class group affects this non-interacting
picture:  A realistic state might be a superposition of position
eigenstates (3.12), with some variance $\sigma^2$.  Thus, the
support of each term in the sum is a fuzzy patch centered at
$\rho(g_i) X$.  If the patches for various mapping class images
overlap, then one expects to pick up interference
terms - this is the essence of ``interactions" in $(2+1)$-dimensional
gravity.  We will define ``in" and ``out" states by demanding that
such overlaps do not occur - practically, this requires that the
$X(\mu )$ be ``sufficiently large", so that for every element
$g_i$ of the mapping class group, the image of $X(\mu )$ under
$g_i$  lies in the tail of the wave packet, which is exponentially
suppressed:

$$\forall \{ \mu, i\} , \parallel X(\mu) - \rho(g_i) X(\mu)
\parallel^2 >> \sigma^2 \eqno(3.13)$$

\noindent This criterion depends on the variance of the wave
packet; one can get a more universal condition if one assumes
that the variance is of order $\hbar^2$.  In units $\hbar = 1$
this gives

$$\forall \{ \mu, i \}, \parallel X(\mu) - \rho (g_i) X (\mu)
\parallel^2 >> 1 \eqno(3.14)$$

	We are now ready to calculate the scattering amplitude:  Taking
two asymptotic invariant states (3.12) and the Hamiltonian (2.24),
and recalling that the mapping class group leaves H invariant and
translates T by a constant, the amplitude is just

$$\eqalign{ < 1 & \vert 2 > \sim \sum_{i,j \in \{m.c.g.\}}
<\rho(g_j) (X_2, T_2) \vert e^{iH(T_2^j - T_1^i)} \vert \rho(g_i)
(X_1, T_1) > \cr}\eqno(3.15)$$

$$\eqalign{ \sim \sum_{i,j \in \{m.c.g.\} } & < X_2, T_2 \vert
e^{-iH(T_2 - T_1^{i-j})} \vert  \rho(g_j{}^{-1}) \rho(g_i)
(X_1, T_1) > \cr}\eqno(3.16)$$

\noindent where $\{ X_2^j, T_2^j \} = \rho (g_j) \{ X_2, T_2 \}$, etc..
Finally, since no loop of $\pi_1(\Sigma)$ is privileged with respect to
any other, the products $(g_j^{-1} g_i)$ will cover the mapping
class group uniformily, and absorbing the infinite factor equal to
the cardinality of the group,

$$<1\vert 2> \sim \sum_{i\epsilon \{m.c.g.\} } <X_2, T_2\vert
e^{-iH(T_2-T_1^i)} \vert\rho(g_i) (X_1, T_1)> \eqno(3.17)$$

Thus, the amplitude at $\{ X_2, T_2 \}$ is a sum over the contributions
of each mapping class image of the source, $\{ X_1^i, T_1^i \}$,
propagated to $\{ X_2, T_2 \}$ by the ``free-particle" Hamiltonian
$H({P(\mu ), \mu =2,..,2g+N})$.  Note that the sum is infinite,
and the expression (3.17) is not necessarily well-defined.  We
will argue below that (3.17) is computable as a distribution on
a certain limited set of wave functions $\psi (\{ X_1, T_1 \} )$; it
would be very interesting to identify an appropriate space of
test functions for which the distribution (3.17) is well-defined, but
this lies beyond the scope of this article - the difficulties
arise because the mapping class group does not act properly
discontinuously on the space of polygons.

\noindent 3.4 The stationary phase theorem, and the sum over
mapping class images

	We will consider an initial wave packet centered
at a suitable $X_1^0$, where by ``suitable" we mean that the images
$\{ X_1^i, T_1^i \}$ have no accumulation point
$( \exists \epsilon : \forall i, \sum_{\mu} (X(\mu) -
\rho(g_i) X(\mu))^2 > \epsilon)$.  The orientation of the ``beam" is
defined in a frame-independent way, to get an $SO(2,1)$-invariant
wave function.  We further require that the wave packet a(K) have
support only over $P$'s which correspond to solutions of the cycle
conditions which form a faithful representation of the fundamental
group  [5],[21] (other sheets of solutions include totally
collapsed handles, or curvature singularities with a surplus angle
equal to a multiple of $4 \pi$).

$$\psi (X_1, T_1) = \int dKa(K) e^{iK\cdot (X_1 - X^0_1)}
\eqno(3.18)$$

\noindent where

$$K = \{P^a(\mu); \mu = 2,3, \cdots , 2g+N\} \eqno(3.19)$$

$$dK = \Pi_\mu \ d^3 P(\mu) \eqno(3.20)$$

	The wave function at $(X_2, T_2)$ is given by propagating the
initial state (3.18) with the invariant propagator (3.17).  With
the notation $\omega_k = H(K)$ and the definition

$$\{ X_{(i)}, T_{(i)} \} = \rho(g_i) (X_1, T_1) - (X_2, T_2)
\eqno(3.21)$$

\noindent the wave function becomes

$$\psi(X_2, T_2) \sim \sum_{i\epsilon \{m.c.g.\} } \int dK a(K)e^{i(K
\cdot X_{(i)} - \omega_k T_{(i)}) } \eqno(3.22)$$

	The stationnary phase argument implies that the only contributions
to the integral are from values of $K$ where the phase
varies slowly as compared to $a(K)$; this picks out
the trajectories near the classical solutions

$$X_{(i)} = {\partial \omega_k\over \partial K} T_{(i)} \eqno(3.23)$$

\noindent where the first factor in the r.h.s. is the group velocity
of the packet, which can be calculated explicitly given the Hamiltonian
(2.24).  In the scattering problem, the initial state, final state
and $i$ are fixed, so that (3.23) can be solved for $K$; let us
denote this solution by $K(i)$.  The only elements of the
mapping class group which give a large contribution to the sum (3.22)
are such that $K(i)$ is not in the exponentially suppressed tail
of the wave packet: $a(K(i)) > \epsilon$ for some appropriately
chosen $\epsilon$. Since we have assumed
that the images of the source have no accumulation point, one
expects that only a finite number of terms will contribute and the
wave function (3.22) has a finite norm.  This
is necessarily a rough statement at this point; a general proof
of computability of the expression (3.22) for some appropriate
Hilbert space of wave functions is not yet available.  We have
examined a number of specific examples to various degrees of
completeness, and found that the expression (3.22) is computable
for initial data peaked about sufficiently non-singular polygons.
One encounters the following situations.

\noindent (1)  If the images are widely spaced and their
emissions do not interfere, only one of the classical trajectories
$K(i)$ lies within the wave packet $a(K)$; one recovers the
free propagator.  This occurs when the images $X(i)$ are sufficiently
distant from each other as compared to the distance between $X_2$ and
$X_1$, so that the images $K(j)$ different from
$K(i)$ do not lie within the support of the wave packet
$a(K)$.  There is no scattering in this case.

\noindent (2)  Two classical trajectories connect the initial
and final states, such as with geodesics on a cone (two-particle
scattering).   If the initial wave packet is wide enough, e.g.
includes both $P(1)$ and $M(2)P(1)$, then the scattering amplitude
involves the interference between these two terms, as in a two
slit experiment.  Other terms in the sum, such as $(M(2))^n P(1)$
for $n>1$, correspond to the events where one particle winds
around the other $n$ times - the corresponding $K$ may lie within
the packet $a(K)$ for small enough $n$, but these trajectories
are not ``nearly classical" and will contribute little, by the
stationnary phase agument.  According to 't Hooft's more
complete analysis [13], such terms are suppressed by a factor of order
$(\epsilon / 2 \pi)^{4n}$, where $\epsilon$ is the deficit angle
at the conical singularity.

\noindent (3)  One of the braid generators of the mapping class
group (winding around a small loop) acts as a small translation,
so that one picks up interference from a large number of images
spaced on a line:  This amplitude is related to scattering of
light by a grating, and describes the quantum dynamics of a thin
wormhole. If one of the mapping
class generators $g_{\alpha}$ modifies $K(i)$ very little, where
$K(i)$ is a classical solution which lies within the packet a(k),
then by applying this generator again and again one
obtains a sequence of contributions from $K_{(j_n)}$, with $g_{(j_n)} =
(g_{\alpha})^{\ n} g(i)$.  The series (3.22) may be exactly summable
in this case; in the example of the torus the result should be
related to the Maass forms [16].

\noindent (4)  Various mapping class generators contribute
significantly to the sum:  In the general case where more than
one generator has ``almost non-proper" action on $K$, one must
consider the interference of the amplitudes propagating freely
from a large number of images, which are distributed in a
complicated fashion. This corresponds to a multiple
scattering amplitude, and the expression (3.22) is related to
the scattering by a complicated multidimensional ``grating".

\

\noindent 3.5 Quantum gravity near the ``big bang", and through
the ``big bounce"?

	Besides the scattering from asymptotically free states,  another
problem of interest is the $(2+1)$-dimensional equivalent of
the ``big bang".  Many solutions of the classical theory expand
from an initial singularity (although smoothly, unlike in $3+1$
dimensions).  For $N=0$ and genus $g \geq 2$,  all  solutions
have either an initial or a final singularity [27].  This brings
up a very interesting situation:  As one approaches the singularity,
the topological features are increasingly tiny and a wave packet of
given variance will contain an increasing amount of mapping class
images.  This means that the number of terms in the invariant wave
function (3.22) increases, and it becomes less and less reasonable
to interpret the result in the semiclassical picture, by referring
to a background geometry with
interference terms acting as effective interactions.  The ``geometry"
becomes completely fuzzy, but the wave function can still be
calculated - this gives one a window on the non-perturbative regime,
and small scale structure, of a theory of quantum geometry.

	The wave function (3.22) is probably computable up to any desired
level of accuracy at a fixed point away from the singularity,
although the number of terms which must be
considered in the sum can be large if one is close to the
initial singularity.  Short of actually performing this
calculation, one can draw a few tentative conclusions.  The
increasing density of mapping class images as one approaches
the big bang indicates that the inevitable anisotropy of classical
universes with non-trivial topology is likely to be smoothed
out:  Any effect which occurs before the quantum  gravity effects
lose significance is likely to appear isotropic to a future
observer, even in a topologically non-trivial cosmology.  By
the same argument, it appears plausible that the wave function at
$\{ X, T \}$ away from the big bang would depend very little on
the precise initial condition on $\psi (X_1, T_1)$.  This would
be very convenient, since it eliminates the problem of having to specify
an initial condition for the very early universe; it places all
the burden of selecting among the possible histories, on the
measurements which are presumably performed later on, and
collapse the wave function to something resembling a classical
background geometry.

	The last compact universe problem which we wish to discuss is
the ``big bounce". The initial state is a wave packet centered
about a contracting universe, which would classically contract to a
singularity and then expand again on the other side.  The
semiclassical picture breaks down
as one approaches the singularity, since the wave function
becomes a superposition of an increasingly large number of
mapping class images and the wave function extends to cover
all possible universes with the given topology.
It is impossible to think in terms of a ``geometry" in this region,
yet interference patterns from the mapping class images is likely
to create ``dark regions" and ``bright regions" in the space of
geometries.  It appears that the large number of mapping class
images may lead to randomization of the wave packet, and a very
non-localized wave function on the other side, as in the previous case.
This may be related to the chaotic behavior of particles close to
the singularity [28].

\

\noindent 3.6 Scattering of Particles and Wormholes on $\Real ^{2}$

	If the universe has the topology $\Real^2$ with $g$ handles and
$N$ punctures, the amplitudes are calculated in much the same way
but the Hamiltonian function is simpler.  The geometry at
infinityis that of a cone with a helical shift, and the axis of
the cone defines the direction of the ``time" vector.  The deficit
angle of the conical geometry gives the total energy, $H$, and the
helical shift corresponds to the total angular momentum [29] [11].  The
cycle condition for closed topologies is replaced by a single
equation, which gives the overall holonomy for a loop which
goes around the circle at infinity; we will consider only wave
functions with support limited by the condition that this overall
holonomy is a rotation, since otherwise closed timelike curves
occur in the semiclassical region (this implies that there is at
least one particle, since wormholes are always tachyonic [21]).
The angle of this rotation is the Hamiltonian:

$$ H = Arc sin (P) \eqno(3.24)$$

\noindent where

$$P^a = {1\over 2} \epsilon^{abc} \biggl( M(1)M^{-1}(2)M^{-1}(1)M(2)
\cdots M(2g+N)\biggr)_{cb} \eqno(3.25)$$

	This function has been written down explicitly as a function of
the momenta for two particles [11] and three particles [14], and
can be calculated, with an increasing amount of work, for any
number of particles.  Note that there is an ambiguity in the
definition of the Hamiltonian, which is only given modulo
$2 \pi$.  This ambiguity is related to the fact that the holonomies
are representations of the fundamental group in the covering
group of $SO(2,1)$.  It is usual to restrict one's attention
to the case where the holonomy $M(\mu )$ associated to each
individual particle ($\mu =2g+1, ..., 2g+N$) lies on the
component simply connected to the identity, and similarly for
the overall holonomy at infinity - this corresponds to particles
of reasonably small mass [21], and so-called ``physically reasonable"
conditions at infinity [30]; the consequence of these asumptions
is that for the $g=0$ case, one can show that there exists a
global spacelike slicing (no time machines) [31]; this fact
was first stated without proof in [11].  We will enforce
these conditions by choosing wave packets with support limited
to these ``reasonable geometries", for $g=0$.  In the general
case, there is no avoiding this ambiguity, and one must decide
if the wave function is multivalued or if it transforms
non-trivially under a $2 \pi$ rotation [23].  As before, we will
make the choice of a scalar wave function and postulate

$$\Psi(X,T) \sim \sum_{i\in \{m.c.g.\} } \int dK a(K) e^{i(K\cdot
X_{(i)} - \omega_k T_{(i)}) } \eqno(3.26)$$

	This expression is similar to Carlip's proposal for exact
scattering amplitudes [23], the only difference being that we
are considering wave packets, rather than the pure propagator.
This allows us to appeal to the stationnary phase theorem and
argue that the sum is limited to a finite number of mapping
class images, which are such that the classical trajectory which
connects an image to $\{ X, T\}$ corresponds to a momentum
$K(i)$ which lies within the support of the packet $a(K)$.  It
would be very interesting to see an explicit solution of the
three-particle scattering problem in a specific situation, by
using (3.28) and the expanded expression for the Hamiltonian [14].

\vfill
\eject

\section{4. Conclusion; the ``problem of time"}

 In this article, we have developped the covariant
quantization programme which
was initiated by Witten and Carlip, using the ``polygon
representation" of the reduced phase space.  In this representation,
the classical $(2+1)$-dimensional gravity appears formally similar
to a set of free particles propagating in $\Real^{2+1}$. We
gave the explicit expression for the Hamiltonian function
and quantized the theory canonically. The mapping-class
invariant wave function can be written as a sum of
freely propagating amplitudes, where each term represents the
propagation from one mapping-class image of the initial state.
Interactions occur when these amplitudes interfere.  This can
produce interference patterns analogous to either the two-slit
experiment, or the diffraction by a regular grating, in some special
cases.

 It is not obvious that the sum over mapping class images is
computable for wave functions which have support over singular
geometries. Since singular geometries occur rather generically
in $(2+1)$-dimensional gravity, it would be of great interest
to prove that the sum is well-defined for a specific Hilbert
space of wave functions.

 Also of interest, is the opportunity to compare this quantum
theory for the torus ($g = 1$), to other author's results in
the extrinsic time.  Joining our results and those of Carlip [9],
we have indirectly shown that the theories are closely related
if one considers only $static$ issues, since it was shown that
the ADM, Witten-Carlip and Polygon variables are different
representations of the same reduced phase space, and that the relations
between these representations are consistent with the operator
orderings chosen [9].  However, it is not clear whether the quantum
dynamical theories, given by Schroedinger's equation in York's time
or in our internal time $T$, are equivalent.  We will argue below
that they are not, but first we must define what we mean
by ``equivalent".  We will use the term ``observable" in the
traditional sense of quantum mechanics, without requiring
that they should be constants of the motion.
An example of an observable might be, e.g., $L^2_{\mu} = E^2(\mu)$.

\noindent $\underline{Definition}$ Two quantum systems, described by
the wave functions $\psi$ and $\psi '$ in times $t$ and $t'$ , are
said to be ``equivalent", if for any set of observables
{$A_i$, $(i = 1, ..., N)$}, the expectation values $<A_i>_t$ and
$<A_i>_{t'}$ follow the same trajectories in $\Real^N$.

\noindent  Let $\{ C_i, H \}$ be a complete set of
commuting observables, where $H$ is the Hamiltonian for time $t$,
and let $K$ be the Hamiltonian for time $t'$.  The wave functions
$\psi$ and $\psi '$ can be expanded in a basis of eigenstates of
the commuting observables.  We now explain why the theories of
$(2+1)$-dimensional gravity in different times are not equivalent.

\noindent (1) K must be diagonal in this basis (if it were not,
then the expectation values of the observables in the state $\psi '$
would not be constant and the equivalence criterion would be violated).

\noindent (2) The initial state $\psi ' (t' = 0)$ must have the
same probability for each CSCO eigenstate as $\psi (t = 0)$
(a trivial consequence of the definition above: set $C_i = A_i$).

\noindent (3) The wave functions $\psi (t)$ and $\psi ' (t')$
differ only by the phase factor of each coefficient in the
expansion in the basis of eigenstates. The time-dependence of
these phases in the two formulations are given by $\phi '_n(t') =
\phi '_n(0) + K_n t'$ and $\phi_n(t) = \phi_n(0) + H_n t$. One easily
checks, by comparing the form of the Hamiltonian for our choice of
internal time $T$ to that which corresponds to the
extrinsic time, that these phases do not follow
the same trajectory in $\Real^{d}$, where $d$ is the number of
commuting observables.  This further implies that one can easily
find a set of $N=2$ observables (which do not commute with $H$)
such that their expectation values do not follow the same
trajectory in $\Real^2$. Thus, either the quantum theories
in different times give different dynamics, or they cannot be
compared in the sense that it is impossible to set up the
same experiment in both pictures. The attempts at avoiding this
fundamental problem by appealing to the path integral formulation
of quantum mechanics, as a unified formalism to which both
Schroedinger quantizations are presumably equivalent, usually
end up hiding the problem in either an inadequate definition of the
boundary conditions, or of the measure: as Guven and Ryan have
emphasized [34], two inequivalent theories cannot be simultaneously
equivalent to a third one!

 Until one can gain a better grasp on this problem, and find an
appropriate Hilbert space for which the wave function (3.22) is
computable, it is impossible to claim that this canonical
formalism for $(2+1)$-dimensional gravity is acceptable as such;
probably it will be necessary to look first at the
simplest case $g = 1$, where it is possible to compare the
results with those of the ADM school.

 One might consider extending this work to $3+1$ dimensions along the
lines of the exactly solvable theories proposed by Horowitz [35].
A reduced phase space for Horowitz's topological $B \wedge
F$ theory can be found [36, 37], and it is probable that the
canonical quantization programme can be carried out along the same
lines as those laid out in this article.

 The canonical quantization programme of $(2+1)$-dimensional
gravity was faced with two problems: that of computing
the Hamiltonian, without which quantum dynamical issues
cannot be addressed, and that of finding wave functions
that are invariant under the action of the
mapping class group. The first problem has essentially been
solved, although this required abandoning the extrinsic gauge
which is favored by the ADM school for its relevance in $3+1$
dimensions; as we argued above, the change of internal time
is not a trivial step in a quantum theory!  As for the second
problem, an explicit construction of the invariant wave
functions was given. Even if the computability cannot
be established in the general case, we hope that this article
has provided a practical scheme with which one can tackle specific
non-perturbative problems of quantum gravity in $2+1$ dimensions.

\

\noindent Acknowledgements

	I would like to express my gratitude to the Institute
for Theoretical Physics (ITP) of the University of California at
Santa Barbara, for their invitation to participate in the workshop
on the small scale structure of spacetime, as well as the Facultad
de Astronom\ii a, Matem\' aticas y F\ii sica of the University of
C\' ordoba, Argentina, the Facultad de Ciencias Exactas y Naturales and
the Instituto de Astronom\ii a y F\ii sica del Espacio (IAFE) of the
University of Buenos Aires, and the Facultad de Ciencias Exactas and
the Instituto de F\ii sica Rosario (ifir) of the National University of
Rosario, Argentina, for their wonderful hospitality while much of this
work was being completed.  We are endebted to Luis Urrutia for
enlightening discussions at the early stages of this work.  We
also gladly acknowledge valuable input from Akio Hosoya, Steve
Carlip, Karl Kuchar and Vince Moncrief.

\vfill
\eject

\section{References}
\parindent=1.2pc

\noindent [1]	B. S. DeWitt, Phys. Rev. 160 (1967), 1113

\noindent [2]	T. Jacobson and L. Smolin, Nonperturbative Quantum
Cosmologies, Nucl. Phys. B299 (1988), 295

A. Ashtekar, Canonical quantum gravity,  Proceedings
of the 1990 Banff Summer School on Gravitation, ed. by R. Mann
(World Scientific, Singapore, 1990)

C. Rovelli, Ashtekar formulation of General Relativity
and Loop Space Non-perturbative Quantum Gravity: A Report,  Class.
Quant. Grav. 8 (1991)

\noindent [3]	E. Witten, Commun. Math. Phys. 121 (1989), 351

\noindent [4]	J. D. Brown, Lower-dimensional Gravity  (World
Scientific, 1988)

\noindent [5]	Ed Witten, Nucl. Phys. B311 (1988) 46

A. Achucarro and P. Townsend, Phys. Lett. B180 (1986), 89

\noindent [6]	J. E. Nelson and T. Regge, Commun. Math. Phys. 141
(1991), 211

J. E. Nelson and T. Regge, Phys. Lett. B272 (1991), 213

J. E. Nelson, T. Regge, Nucl. Phys. B328 (1989), 190

\noindent [7]	V. Moncrief, Ann. Phys. (N.Y.) 167 (1986), 118

V. Moncrief, J. Math. Phys. 30 (1989), 2297

\noindent [8]	S. Carlip, Phys. Rev. D42 (1990), 2647

\noindent [9]	S. Carlip, Phys. Rev. D45 (1992), 3584

S. Carlip, The modular group, operator ordering and
time in 2+1 dimensional gravity,  Univ. of California Preprint
UCD-92-23 (gr-qc/9209011)

\parindent=1.4pc

\noindent [10]	R. Arnowitt, S. Deser and C. W. Misner, Gravitation,
and Introduction to Current Research, [L, Witten, ed.]; Wiley N.Y.
(1962)

J. York, Phys. Rev. Lett. 28 (1972), 1082

\noindent [11]	S. Deser, R. Jackiw and G. 't Hooft, Ann. Phys.
(N. Y.) 152 (1984), 220

\noindent [12]	S. Deser and R. Jackiw, Commun. Math. Phys. 118
(1988), 495

\noindent [13]	G. 't Hooft, Commun. Math. Phys. 117 (1988), 685

\noindent [14]	D. Lancaster and N. Sasakura, Class. Quantum Grav. 8
(1991), 1481

\noindent [15]	A. Anderson: ``Unitary equivalence of the metric and
holonomy formulations of 2+1 dimensional quantum gravity on a torus",
Mc Gill University preprint 92-19, Imperial College preprint
TP/92-93/02  (gr-qc/921007).

A. Anderson: ``Quantum canonical transformations:  Physical
equivalence of quantum theories", Imperial College preprint
TP/92-93/20  (hep-th/9302062)

\noindent [16]	H. Maass, Lectures on Modular Functions of One
Complex Variable (Tata Institute, Bombay, 1964)

J. D. Fray, J. Reine Angew. Math. 293 (1977), 143

\noindent [17]	J. E. Nelson, T. Regge, F. Zertuche, Nucl. Phys. B339
(1990), 516

L. F. Urrutia and F. Zertuche, Class. Quant. Grav. 9
(1992), 641

\noindent [18]	L.F. Urrutia, H. Waelbroeck, F. Zertuche,
Mod. Phys. Lett. A7 (1992), 2715

\noindent [19]	H. Waelbroeck and F. Zertuche, ``Time and the
Hamilton Jacobi Transformation in 2+1 Dimensional Gravity",
to appear jointly with this article.

\noindent [20]	J. S. Birman and H. M. Hilden 1971 On the mapping
class groups of closed surfaces as covering spaces Advances in the
Theory of Riemann Surfaces (Ann. Math. Studies 66)  ed L. V.
Ahlfors et al (Princeton, N. J.: Princeton University	Press)

J. S. Birman 1977 The algebraic structure of surface
mapping class groups Discrete Groups and Automorphic Functions
ed W. J. Harvey (New York: Academic)

\noindent [21]	H. Waelbroeck, Nucl. Phys. B364 (1991), 475

\noindent [22]	G. 't Hooft, Class. Quantum Grav. 9 (1992), 1335,
ibid., to be published (1993)

G. 't Hooft, Canonical Quantization of Gravitating Point
Particles in 2+1 Dimensions, University of Utrecht Preprint THU-93/10

\noindent [23]	S. Carlip, Nucl. Phys. B324 (1989), 106

\noindent [24]	H. Poincar\' e, Acta Mathematica 1 (1882), 1

J. Milnor, Adv. Math. 25 (1977), 178

\noindent [25]	S. Carlip, Class. Quantum Grav.8 (1991), 5

\noindent [26]	A. Hosoya and K. Nakao, Class. Quant. Grav. 7 (1990), 63

\noindent [27]	G. Mess, Lorentz spacetime of constant curvature,
Institut des Hautes Etudes Scientifiques Preprint IHES/M/90/28 (1990)

\noindent [28]	G. 't Hooft, private communication

\noindent [29]	M. Henneaux, Phys. Rev. D29 (1984), 2766

\noindent [30]	S. Deser, R. Jackiw and G. 't Hooft, Phys. Rev. Letters 68
(1992), 267

\noindent [31]	A. Guth et al. M.I.T. Preprint (1992)

\noindent [33]	R. Puzio, ``On the square root of the Laplace-Beltrami
operator as a Hamiltonian", Yale University Preprint, Sep. 13 1993.

\noindent [34]	J. Guven and M. Ryan, Phys. Rev. D45 (1992),

\noindent [35]	G. T. Horowitz, Commun. Math. Phys. 125 (1989), 417

\noindent [36]	H. Waelbroeck , `` $B \wedge F$ theory and flat
spacetimes", Mexico preprint ICN-UNAM-93-12 (gr-qc/9311033)

\vfill
\eject

\noindent [37] H. Waelbroeck and J. A. Zapata, ``A Hamiltonian
Formulation of Topological Gravity", to appear in Class. Quant. Grav.

\

\

\

\section{Figure Captions}

\noindent $\underline{Figure \ 3.1}$  The loop $a_1 b_1$ crosses twice
through the edges of the original polygon, at $X=X'$ and at $Y=Y'$.
By recutting the surface in a different way prior to unwrapping
it as a polygon, one obtains a first cut which goes directly from
$A$ to $C$ (dotted line), so that the loop $a_1 b_1$ crosses only
once an edge of the new polygon, exactly in the same way as $a_1$
crossed one edge of the original polygon.  The switch from one
set of cuts to another, and from $a_1$ to $a_1 b_1$, is an example of
a mapping class transformation.

\end